\pdfoutput=1
\documentclass[twocolumn,physrev,superscriptaddress]{revtex4-2}
\usepackage{amsmath,amssymb}
\usepackage{graphicx}
\usepackage{epstopdf}
\usepackage{color}
\usepackage{paralist} 
\usepackage{lineno} 

\usepackage{placeins}
\usepackage{enumerate}
\usepackage{oubraces}
\usepackage{tikz}
\usetikzlibrary{shapes.geometric}
\tikzset{gauge1/.style={draw=none,minimum size=0.4cm,fill=white,circle, draw}}
\tikzset{gauge5/.style={draw=none,minimum size=0.6cm,fill=white,circle, draw}}
\tikzset{supergauge/.style={draw=none,minimum size=0.9cm,fill=white,circle, draw}}
\tikzset{bluenode/.style={draw=none,minimum size=0.4cm,fill=blue,circle, draw}}
\tikzset{bluegauge/.style={draw=none,minimum size=0.4cm,fill=blue,circle, draw}}
\tikzset{rednode/.style={draw=none,minimum size=0.4cm,fill=red,circle, draw}}
\tikzset{redgauge/.style={draw=none,minimum size=0.4cm,fill=red,circle, draw}}
\tikzset{gauge3/.style={draw=none,minimum size=0.4cm,fill=white,circle, draw}}
\tikzset{dotsize/.style={draw=none,minimum size=0.6pt,fill=black,circle,inner sep=1pt, draw}}
\tikzset{mini/.style={draw=none,minimum size=1pt,fill=white,circle,inner sep=3pt, draw}}
\tikzset{miniG/.style={draw=none,minimum size=1pt,fill=black,circle,inner sep=3pt, draw}}
\tikzset{cyane/.style={draw=none,minimum size=0.4cm,fill=cyan,circle, draw}}
\tikzset{pinklinet/.style={draw=none,minimum size=0.4cm,fill=magenta,circle, draw}}
\tikzset{greenlinet/.style={draw=none,minimum size=0.4cm,fill=green,circle, draw}}
\tikzset{blacknode/.style={draw=none,minimum size=0.4cm,fill=black,circle, draw}}
\tikzset{brownlinet/.style={draw=none,minimum size=0.4cm,fill=olive,circle, draw}}
\tikzset{magicmintlinet/.style={draw=none,minimum size=0.4cm,fill=red,circle, draw}}
\tikzset{orangeet/.style={draw=none,minimum size=0.4cm,fill=orange,circle, draw}}
\tikzset{grayet/.style={draw=none,minimum size=0.4cm,fill=gray,circle, draw}}
\tikzset{blueet/.style={draw=none,minimum size=0.4cm,fill=blue,circle, draw}}
\tikzset{flavour2/.style={draw=none,minimum size=0.8cm,fill=white, regular polygon,regular polygon sides=4,draw}}
\tikzset{flavour2/.style={draw=none,minimum size=0.6cm,fill=white, regular polygon,regular polygon sides=4,draw}}
\tikzset{redflavor/.style={draw=none,minimum size=0.6cm,fill=red, regular polygon,regular polygon sides=4,draw}}
\tikzset{redsquare/.style={draw=none,minimum size=0.6cm,fill=red, regular polygon,regular polygon sides=4,draw}}
\tikzset{bluesquare/.style={draw=none,minimum size=0.6cm,fill=blue, regular polygon,regular polygon sides=4,draw}}
\tikzset{greenflavor/.style={draw=none,minimum size=0.6cm,fill=green, regular polygon,regular polygon sides=4,draw}}
\tikzset{brownflavor/.style={draw=none,minimum size=0.6cm,fill=brown, regular polygon,regular polygon sides=4,draw}}
\tikzset{pinkflavor/.style={draw=none,minimum size=0.6cm,fill=magenta, regular polygon,regular polygon sides=4,draw}}
\tikzset{grayflavor/.style={draw=none,minimum size=0.6cm,fill=gray, regular polygon,regular polygon sides=4,draw}}
\tikzset{none/.style={draw=none}}
\tikzset{new edge style 1/.style={dashed}}
\tikzset{dashedline/.style={dashed}}
\tikzset{brace1/.style={decorate,decoration={brace,amplitude=5pt,mirror}}}
\tikzset{bluee/.style={line width=0.5mm,blue}}
\tikzset{orangee/.style={line width=0.5mm,orange}}
\tikzset{magentae/.style={line width=0.5mm,magenta}}
\tikzset{rede/.style={line width=0.5mm,red}}
\tikzset{thickred/.style={line width=5mm,red}}
\tikzset{greene/.style={line width=0.5mm,green}}
\tikzset{darke/.style={line width=0.5mm,black}}
\tikzset{cyaneX/.style={line width=0.5mm,cyan}}
\tikzset{new edge style 3/.style={dashed,red}}
\tikzset{magicmintline/.style={line width=0.5mm,gray}}
\tikzset{brownline/.style={line width=0.5mm,brown}}
\tikzset{greenline/.style={line width=0.5mm,green}}
\tikzset{oliveline/.style={line width=0.5mm,green}}
\tikzset{darkgreenline/.style={line width=0.5mm,olive}}
\tikzset{pinkline/.style={line width=0.5mm,magenta}}
\tikzset{dottedz/.style={line width=0.5mm,black,dotted}}
\tikzset{pinkline2/.style={line width=0.5mm,magenta,dotted}}
\tikzset{brace2/.style={decorate,decoration={brace,amplitude=5pt}}}
\tikzset{reddotted/.style={line width=0.5mm,red,dotted}}
\tikzset{bluedotted/.style={line width=0.5mm,blue,dotted}}
\tikzset{magicmintdotted/.style={line width=0.5mm,gray,dotted}}
\tikzset{greendotted/.style={line width=0.5mm,green,dotted}}
\tikzset{browndotted/.style={line width=0.5mm,brown,dotted}}
\tikzset{arrowed/.style={line width=0.5mm,->}}
\pgfdeclarelayer{edgelayer}
\pgfdeclarelayer{nodelayer}
\usetikzlibrary{decorations.pathreplacing}
\pgfsetlayers{edgelayer,nodelayer,main}

\usepackage{xcolor,colortbl}

\usepackage{ytableau}
\tikzstyle{brane}=[draw]
\tikzset{D7/.style={circle, draw=black, inner sep=0pt, fill=white, minimum size=3mm}}
\tikzset{hasse/.style={circle, fill,inner sep=2pt}}
\tikzset{flavor/.style={regular polygon,regular polygon sides=4,inner sep=2.5pt, draw}}
\tikzset{gauge/.style={circle, draw,inner sep=2.5pt}}
\tikzset{gaugeb/.style={circle, draw,fill=black,inner sep=2.5pt}}
\tikzset{gaugered/.style={circle, draw,fill=red,inner sep=2.5pt}}
\tikzset{gaugeblue/.style={circle, draw,fill=blue,inner sep=2.5pt}}
\tikzset{gaugegreen/.style={circle, draw,fill=green,inner sep=2.5pt}}
\tikzset{bd/.style={circle, draw=black, inner sep=0pt, fill=black, minimum size=2mm}}
\tikzset{wd/.style={circle, draw=black, inner sep=0pt, fill=white, minimum size=2mm}}
\tikzset{Dynkin/.style={circle, draw=black, inner sep=0pt, fill=white, minimum size=2mm}}
\tikzstyle{ligne}=[draw, thick] 
\tikzset{doublearrow/.style={ draw=black!75, color=black!75, thick, double distance=3pt, }} 
\tikzset{bluedouble/.style={ draw=blue!75, color=blue!75, thick, double distance=3pt, }}

\tikzset{reddashed/.style={ draw=red, color=red, thick }}
\tikzset{redarrow/.style={ draw=red, -> , thick }}

\usepackage{enumitem}

\usetikzlibrary{calc}
\tikzset{gaugeJ/.style={inner sep=1mm,draw=none,fill=white,minimum size=2mm,circle, draw}}
\tikzset{flavourJ/.style={draw=none,minimum size=0.3mm,fill=white, regular polygon,regular polygon sides=4,draw}}
\tikzset{hasseJ/.style={circle, fill,inner sep=2pt}}
\setlist{nolistsep}
\usepackage{multirow}

\usepackage[normalem]{ulem}

\begin{document}
\title{An exceptionally simple family of Orthosymplectic 3d $\mathcal{N}=4$ $\text{rank}\scriptsize{-}0$ SCFTs}
\author{Zhenghao Zhong}
\affiliation{Mathematical Institute, University of Oxford, 
Andrew Wiles Building, Woodstock Road, Oxford, OX2 6GG, UK}

\begin{abstract}
\noindent 
We look at a family of 3d $\mathcal{N}=4$ $\text{rank}\scriptsize{-}0$ orthosymplectic quiver gauge theories. We define a superconformal field theory (SCFT) to be $\text{rank}\scriptsize{-}0$ if either the Higgs branch or Coulomb branch is trivial. This family of non-linear orthosymplectic quivers has Coulomb branches that can be factorized into products of known moduli spaces. More importantly, the Higgs branches are all trivial. Consequently, the full moduli space of the smallest member is simply $\mathrm{(one-}F_4 \; \mathrm{instanton}) \times \mathrm{(one-}F_4 \; \mathrm{instanton})$. Although the $3d$ mirror is non-Lagrangian, it can be understood through the gauging of topological symmetries of Lagrangian theories.  Since the 3d mirror possesses a trivial Coulomb branch, we discuss some implications for $\text{rank}\scriptsize{-}0$ 4d $\mathcal{N}=2$ SCFTs and symplectic duality. 
\end{abstract}

\maketitle

\section{Introduction}
Vacuum configurations of quantum field theories are parameterized by vacuum expectation values (VEVs) of scalar fields. In the standard model, this will be the Higgs boson. In supersymmetric gauge theories, the many scalar fields from supermultiplets give rise to a very rich vacuum structure --- moduli space of vacua. For three-dimensional superconformal field theories (SCFTs) with eight supercharges ($3d$ $\mathcal{N}=4$), the supermultiplets are hypermultiplets and vector multiplets. If the moduli space is parameterized by VEVs of scalar fields from the hypermultiplets, it is called the \emph{Higgs branch}. Similarly, if the moduli space is parameterized by scalars in the vector multiplet, it is called the \emph{Coulomb branch}. If there are non-zero VEVs from scalars in both multiplets, we will have a mixed branch. As a result, the full moduli space including all three branches is often a very complicated object. An important infrared (IR) duality for $3d$ $\mathcal{N}=4$ SCFTs, known as 3d mirror symmetry, exchanges the Higgs and Coulomb branches \cite{Intriligator:1996ex}.

Classification schemes for superconformal field theories (SCFTs) have been developed across various space-time dimensions, primarily based on the rank (or Coulomb branch dimension) of the theory. The effectiveness of these classifications often relies on the assumption that interacting SCFTs with rank zero do not exist \cite{Argyres:2020nrr}. For dimensions $d=4,5,6$, it is widely conjectured that, while Higgs branches of many SCFTs can be trivial, there are no interacting SCFTs with a trivial Coulomb branch.  Interacting SCFTs excludes free theories and discrete quotients thereof. However, this assumption has been challenged (for example, discussions on rank-zero $5d$ $\mathcal{N}=1$ SCFTs can be found in \cite{Closset:2020scj}).

Recently, a series of papers \cite{zero,zero11,zero10,zero9,zero8,zero7,zero6,zero5,zero4,zero3,zero2,zero1} has presented compelling evidence for the existence of rank-zero $3d$ $\mathcal{N}=4$ interacting SCFTs. In these works, the definition of rank-zero is more stringent, referring to both the Coulomb and Higgs branches being trivial. Despite having trivial Higgs and Coulomb branches, the $3d$ superconformal index for these SCFTs remains non-zero, supporting the existence of an interacting SCFT. In this paper, we adopt a similar definition as in higher-dimensional cases, where rank-zero requires only one trivial branch. 

For the first time, we present $3d$ $\mathcal{N}=4$ SCFTs with a trivial Higgs branch and a non-trivial Coulomb branch. These are orthosymplectic quiver gauge theories --- containing a combination of special orthogonal and symplectic gauge groups. In fact, it has the following desirable properties:
\vspace{0.05cm}
\begin{itemize}
\setlength{\itemsep}{0pt}
    \item Trivial Higgs branch and non-trivial Coulomb branch. \vspace{0.01cm}
    \item UV theory has a Lagrangian description.  \vspace{0.1cm}
    \item Coulomb branch is a hyperK\"ahler cone.  \vspace{0.1cm}
    \item Complete Higgsing (gauge symmetry can be completely broken by giving VEV to scalars in hypers). 
\end{itemize}

Through $3d$ mirror symmetry, we then construct a theory with a trivial Coulomb branch and a non-trivial Higgs branch. The mirror theory is inherently non-Lagrangian, yet we can represent it as a familiar Lagrangian SCFT with certain topological symmetries gauged. We discuss implications for $4d$ $\mathcal{N}=2$ rank-zero SCFTs.



In Section \ref{THEquiver}, we examine the smallest member of the orthosymplectic family, which has a Coulomb branch isometry of $F_4 \times F_4$, along with its $3d$ mirror. Section \ref{THEgeneralquiver} extends this to an infinite family and discuss how balancing conditions can imply trivial Higgs branches. In Section \ref{THEimplications}, we discuss the implications of this rank-zero family for $4d$ SCFTs and symplectic duality.

\section{Rank-zero orthosymplectic quiver}\label{THEquiver}
The rank-zero family of orthosymplectic quivers previously appeared in \cite{product}, but only their Coulomb branches were studied, with no computation of their Higgs branches. As a result, their significance as rank-zero SCFTs was not recognized. The simplest member of this family is:
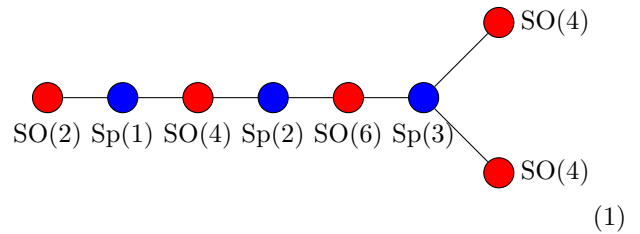
\begin{equation}
\begin{tikzpicture}
	\begin{pgfonlayer}{nodelayer}
		\node [style=redgauge] (0) at (-3, -1) {};
		\node [style=redgauge] (1) at (-1, -1) {};
		\node [style=redgauge] (2) at (1, -1) {};
		\node [style=redgauge] (3) at (3, 0) {};
		\node [style=redgauge] (4) at (3, -2) {};
		\node [style=bluegauge] (5) at (-2, -1) {};
		\node [style=bluegauge] (6) at (0, -1) {};
		\node [style=bluegauge] (7) at (2, -1) {};
		\node [style=none] (8) at (-3, -1.5) {SO(2)};
		\node [style=none] (9) at (-2, -1.5) {Sp(1)};
		\node [style=none] (10) at (-1, -1.5) {SO(4)};
		\node [style=none] (11) at (0, -1.5) {Sp(2)};
		\node [style=none] (12) at (1, -1.5) {SO(6)};
		\node [style=none] (13) at (2, -1.5) {Sp(3)};
		\node [style=none] (14) at (3.75, 0) {SO(4)};
		\node [style=none] (15) at (3.75, -2) {SO(4)};
	\end{pgfonlayer}
	\begin{pgfonlayer}{edgelayer}
		\draw (0) to (7);
		\draw (7) to (3);
		\draw (4) to (7);
	\end{pgfonlayer}
\end{tikzpicture}
\label{quiver}
\end{equation}
This is a quiver diagram where circles represent gauge groups, and lines indicate hypermultiplets transforming in the vector/fundamental representation of the nodes they connect. We use the convention that red nodes represent special orthogonal gauge groups, while blue nodes represent symplectic gauge groups. The gauge groups are explicitly labeled (rather than just their algebras), as the global form of gauge groups affects the moduli spaces non-trivially. This quiver is an example of a non-linear, flavorless orthosymplectic quiver that, due to computational challenges, has only recently been systematically studied in the literature \cite{product,exo2}.

The Coulomb branch of this quiver can be studied by computing its Hilbert series using the monopole formula \cite{Cremonesi:2013lqa}. This theory has a non-trivial one-form symmetry, $\mathbb{Z}_2$, which acts trivially on the matter representations. As a result, there is a choice of whether to gauge this one-form symmetry, which alters the lattice of dressed monopole operators being summed over and, consequently, the Coulomb branch itself \cite{Bourget:2020xdz}. If one chooses to gauge this one-form symmetry, the resulting Coulomb branch is $\mathrm{(one-} F_4 \; \mathrm{instanton}) \times \mathrm{(one-} F_4 \; \mathrm{instanton})$ \cite{product}.

Alternatively, one can choose not to gauge this $\mathbb{Z}_2$ symmetry, resulting in a product moduli space with an $SO(9) \times SO(9)$ Coulomb branch global symmetry. The exact nature of this moduli space remains unidentified. In the literature, it is often conventional—especially when matching with moduli spaces of higher-dimensional SCFTs—to gauge all one-form symmetries. The reasoning behind this convention is not yet fully understood, though it is explored in more detail in \cite{Nawata:2023rdx}.

The Higgs branch is a hyperKähler quotient which takes the space of constant scalar field configurations with vanishing F-terms and quotienting it by the complexified gauge groups. This process can then be translated into computing a Higgs branch Hilbert series (HS):

\begin{equation}
\begin{split}
    \text{HS} &= \oint d\mu \frac{\text{PE}\left( \parbox[]{6cm}{$\sum\limits_{n=1}^{2}(\chi^{\text{vec}}_{\text{SO(2n)}}\chi^{\text{fund}}_{\text{Sp(n)}}t +\chi^{\text{vec}}_{\text{SO(2n+2)}}\chi^{\text{fund}}_{\text{Sp(n)}}t)
    \\ +\chi^{\text{vec}}_{\text{SO(6)}}\chi^{\text{fund}}_{\text{Sp(3)}}t +\sum\limits_{i=1,2}\chi^{\text{vec}}_{\text{SO(4)}_i}\chi^{\text{fund}}_{\text{Sp(3)}}t  $}\right)}{\text{PE}\left( \parbox[]{6.1cm}{$\sum\limits_{n=1}^{3}(\chi^{\text{adj}}_{\text{SO(2n)}}t^2+\chi^{\text{adj}}_{\text{Sp(n)}}t^2 ) +\sum\limits_{i=1,2}\chi^{\text{adj}}_{\text{SO(4)}_i}t^2$}\right)} \\
    & =1
    \label{HSeqn}
    \end{split}
\end{equation}
where $d\mu$ is the Haar measure of the groups, $\chi^{\text{vec}},\chi^{\text{fund}},\chi^{\text{adj}}$ are the characters of the vector, fundamental and adjoint representations, PE is the plethystic exponential and $t$ is the counting fugacity. $SO(4)_{1,2}$ labels the two nodes in the bifurcation. The result $\text{HS}=1$ is an explicit check that the Higgs branch is trivial. The choice of whether to gauge the $\mathbb{Z}_2$ one-form symmetry does not affect the Higgs branch.

\subsection{3d mirror pair}
\begin{figure*}
    \centering
\scalebox{0.7}{\begin{tikzpicture}
	\begin{pgfonlayer}{nodelayer}
		\node [style=redgauge] (0) at (-3, -1) {};
		\node [style=redgauge] (1) at (-1, -1) {};
		\node [style=redgauge] (2) at (1, -1) {};
		\node [style=redgauge] (3) at (3, 0) {};
		\node [style=redgauge] (4) at (3, -2) {};
		\node [style=bluegauge] (5) at (-2, -1) {};
		\node [style=bluegauge] (6) at (0, -1) {};
		\node [style=bluegauge] (7) at (2, -1) {};
		\node [style=none] (8) at (-3, -1.5) {SO(2)};
		\node [style=none] (9) at (-2, -1.5) {Sp(1)};
		\node [style=none] (10) at (-1, -1.5) {SO(4)};
		\node [style=none] (11) at (0, -1.5) {Sp(2)};
		\node [style=none] (12) at (1, -1.5) {SO(6)};
		\node [style=none] (13) at (2, -1.5) {Sp(3)};
		\node [style=none] (14) at (3.75, 0) {SO(4)};
		\node [style=none] (15) at (3.75, -2) {SO(4)};
		\node [style=redgauge] (16) at (-3, 4.25) {};
		\node [style=redgauge] (17) at (-1, 4.25) {};
		\node [style=redgauge] (18) at (1, 4.25) {};
		\node [style=bluegauge] (21) at (-2, 4.25) {};
		\node [style=bluegauge] (22) at (0, 4.25) {};
		\node [style=bluegauge] (23) at (2, 4.25) {};
		\node [style=none] (24) at (-3, 3.75) {SO(2)};
		\node [style=none] (25) at (-2, 3.75) {Sp(1)};
		\node [style=none] (26) at (-1, 3.75) {SO(4)};
		\node [style=none] (27) at (0, 3.75) {Sp(2)};
		\node [style=none] (28) at (1, 3.75) {SO(6)};
		\node [style=none] (29) at (2, 3.75) {Sp(3)};
		\node [style=redflavor] (30) at (3, 4.25) {};
		\node [style=none] (31) at (3, 3.75) {SO(8)};
		\node [style=none] (32) at (-5, 4.25) {$G_T=SO(8)$};
		\node [style=none] (33) at (-5, 3.75) {$G_F=SO(8)$};
		\node [style=none] (34) at (-5, -0.75) {$G_T=F_4\times F_4$};
		\node [style=none] (35) at (-5, -1.25) {$G_F=\emptyset$};
		\node [style=none] (36) at (0, 2.25) {};
		\node [style=none] (37) at (0, 0.5) {};
		\node [style=none] (38) at (-3.75, 1.5) {Gauge $SO(4)\times SO(4)$ flavor symmetry};
		\node [style=redgauge] (55) at (13, 4.25) {};
		\node [style=redgauge] (56) at (11, 4.25) {};
		\node [style=redgauge] (57) at (9, 4.25) {};
		\node [style=bluegauge] (58) at (12, 4.25) {};
		\node [style=bluegauge] (59) at (10, 4.25) {};
		\node [style=bluegauge] (60) at (8, 4.25) {};
		\node [style=none] (61) at (13, 3.75) {SO(2)};
		\node [style=none] (62) at (12, 3.75) {Sp(1)};
		\node [style=none] (63) at (11, 3.75) {SO(4)};
		\node [style=none] (64) at (10, 3.75) {Sp(2)};
		\node [style=none] (65) at (9, 3.75) {SO(6)};
		\node [style=none] (66) at (8, 3.75) {Sp(3)};
		\node [style=redflavor] (67) at (7, 4.25) {};
		\node [style=none] (68) at (7, 3.75) {SO(8)};
		\node [style=none] (69) at (15, 4.25) {$G_T=SO(8)$};
		\node [style=none] (70) at (15, 3.75) {$G_F=SO(8)$};
		\node [style=none] (71) at (15, -0.75) {$G_T=\emptyset$};
		\node [style=none] (72) at (15, -1.25) {$G_F=F_4\times F_4$};
		\node [style=none] (73) at (10, 2.25) {};
		\node [style=none] (74) at (10, 0.5) {};
		\node [style=none] (75) at (14, 1.5) {Gauge $SO(4)\times SO(4)$ topological symmetry};
		\node [style=none] (76) at (4, 4) {};
		\node [style=none] (77) at (6, 4) {};
		\node [style=none] (78) at (5, 4.25) {3d mirror};
		\node [style=none] (79) at (4, -1) {};
		\node [style=none] (80) at (6, -1) {};
		\node [style=none] (81) at (5, -0.75) {3d mirror};
		\node [style=none] (82) at (10, -1) {Non-Lagrangian theory};
	\end{pgfonlayer}
	\begin{pgfonlayer}{edgelayer}
		\draw (0) to (7);
		\draw (7) to (3);
		\draw (4) to (7);
		\draw (16) to (23);
		\draw (23) to (30);
		\draw [style=arrowed] (36.center) to (37.center);
		\draw (55) to (60);
		\draw (60) to (67);
		\draw [style=arrowed] (73.center) to (74.center);
		\draw [style=arrowed] (76.center) to (77.center);
		\draw [style=arrowed] (77.center) to (76.center);
		\draw [style=arrowed] (79.center) to (80.center);
		\draw [style=arrowed] (80.center) to (79.center);
	\end{pgfonlayer}
\end{tikzpicture}
}
    \caption{We start with the $T[SO(8)]$ quivers at the top of the diagram which is self-dual under 3d mirror symmetry. On the left side we gauge the $SO(4)\times SO(4)\subset SO(8)$ subgroup of the flavor symmetry ($G_F$) to arrive at quiver \eqref{quiver} which has trivial Higgs branch and $\mathrm{(one-}F_4 \; \mathrm{instanton})^2$ Coulomb branch. Equivalently, this means one gauge the $SO(4)\times SO(4) \subset SO(8)$ topological symmetry ($G_T$) on the mirror side, arriving at a non-Lagrangian theory with trivial Coulomb branch and $\mathrm{(one-}F_4 \; \mathrm{instanton})^2$ Higgs branch.}
    \label{map}
\end{figure*}
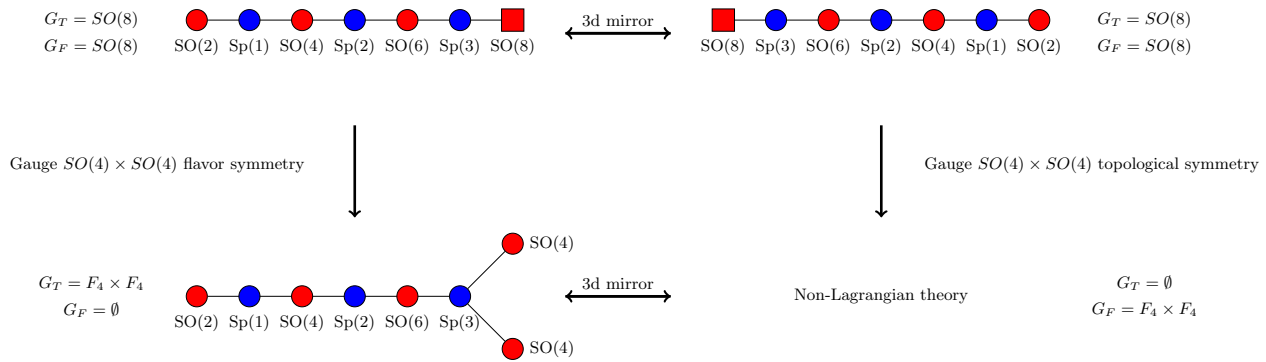

Now, we turn our attention to the 3d mirror theory. Quiver \eqref{quiver} is actually related to a more familiar quiver, $T[SO(8)]$ \cite{Gaiotto:2008ak}. This theory has a Coulomb branch that is the nilcone of $\mathfrak{so}(8)$ (equivalently, the closure of the maximal nilpotent orbit). With an $SO(8)$ flavor symmetry, one can gauge the $SO(4) \times SO(4)$ subgroup to obtain \eqref{quiver}. A unique property of $T[SO(8)]$ is its self-duality under 3d mirror symmetry, meaning its Coulomb branch is identical to its Higgs branch. Given a 3d mirror pair, the Coulomb branch (Higgs branch) of one theory corresponds to the Higgs branch (Coulomb branch) of its mirror. Accordingly, the global symmetry (isometry) groups are also identical: the Coulomb branch global symmetry is the topological symmetry enhanced in the IR by monopole operators, while the Higgs branch global symmetry corresponds to the classical flavor symmetry.
Thus, the $SO(8)$ flavor symmetry of $T[SO(8)]$ is equivalent to the $SO(8)$ topological symmetry of its identical mirror pair. Additionally, gauging the $SO(4) \times SO(4)$ subgroup of the flavor symmetry translates to gauging the $SO(4) \times SO(4)$ topological symmetry in the mirror theory. These results are summarized in Figure \ref{map}.

The 3d mirror theory is a rank-zero theory with a trivial Coulomb branch but Higgs branch of $\mathrm{(one-}F_4 \; \mathrm{instanton})^2$. It is naturally a non-Lagrangian theory as one would expect from a theory with trivial Coulomb branch, nevertheless we can easily construct it from the well known $T[SO(8)]$ theory. Even though we cannot compute the moduli space of this non-Lagrangian theory directly, we are able to find it using mirror symmetry.  Thus, we have constructed quiver \eqref{quiver}, whose full moduli space is its Coulomb branch, and its 3d mirror, whose full moduli space is just its Higgs branch.

\section{Family of rank-zero quivers}\label{THEgeneralquiver}
Quiver \eqref{quiver} can be easily extended by taking the $T[SO(2n)]$ quiver and gauging $SO(n)\times SO(n)\subset SO(2n)$ flavor symmetry:
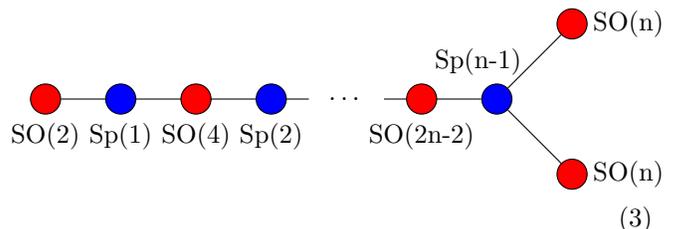
\begin{equation}
\begin{tikzpicture}
	\begin{pgfonlayer}{nodelayer}
		\node [style=redgauge] (0) at (-4, -1) {};
		\node [style=redgauge] (1) at (-2, -1) {};
		\node [style=redgauge] (3) at (3, 0) {};
		\node [style=redgauge] (4) at (3, -2) {};
		\node [style=bluegauge] (5) at (-3, -1) {};
		\node [style=bluegauge] (6) at (-1, -1) {};
		\node [style=bluegauge] (7) at (2, -1) {};
		\node [style=none] (8) at (-4, -1.5) {SO(2)};
		\node [style=none] (9) at (-3, -1.5) {Sp(1)};
		\node [style=none] (10) at (-2, -1.5) {SO(4)};
		\node [style=none] (11) at (-1, -1.5) {Sp(2)};
		\node [style=none] (13) at (1.75, -0.5) {Sp(n-1)};
		\node [style=none] (14) at (3.75, 0) {SO(n)};
		\node [style=none] (15) at (3.75, -2) {SO(n)};
		\node [style=redgauge] (16) at (1, -1) {};
		\node [style=none] (17) at (0.5, -1) {};
		\node [style=none] (18) at (-0.5, -1) {};
		\node [style=none] (19) at (0, -1) {\dots};
		\node [style=none] (20) at (1, -1.5) {SO(2n-2)};
	\end{pgfonlayer}
	\begin{pgfonlayer}{edgelayer}
		\draw (7) to (3);
		\draw (4) to (7);
		\draw (18.center) to (0);
		\draw (17.center) to (7);
	\end{pgfonlayer}
\end{tikzpicture}
\label{generalfamily}
\end{equation}
 For $n=4$, we obtain quiver \eqref{quiver}. For $n\geq 5$, the Coulomb branch global symmetry is $SO(2n+1) \times SO(2n+1)$. The Coulomb branches of this family  are product moduli spaces and given in \cite{product}. For $n\geq 4$, computing the Higgs branch Hilbert series becomes increasingly challenging, though it can still be calculated perturbatively to some extent.

\subsection{Balance and dimension of Higgs branch} 
Another diagnostic for a trivial Higgs branch is to compute the Higgs branch dimension. Since the Higgs branch is a hyperK\"ahler quotient, the Higgs branch dimension is simply:
\begin{equation}
    \text{dim}_{\mathbb{H}} =\frac{1}{2}( \# \text{hypers} - \sum_i\text{dim}(G_i))
    \label{dimformula}
\end{equation}
where $\sum_i\text{dim}(G_i)$ is the sum of the complex dimensions of the gauge groups and the overall 1/2 is to convert complex to quarterionic dimensions. In quiver \eqref{generalfamily}, we observe that all the gauge nodes are balanced, as defined in \cite{Gaiotto:2008ak} for orthosymplectic gauge groups, which means each $SO(2n)$ gauge group is connected to $2n-1$ full hypers, each $SO(2n+1)$ is connected to $2n$ full hypers and each $Sp(n)$ gauge group is connected to $2n+1$ full hypers. The \#hypers of an $SO(k) - Sp(j)$ edge is $\frac{kj}{2}$. We observe here that a quiver made of only balanced gauge groups (and no flavor groups)  has $\text{dim}_{\mathbb{H}}=0$. 

To find quiver gauge theories with trivial Higgs branch, we just have to find flavorless and fully balanced quivers. However, this is easier said than done. Consider the affine-Dynkin quiver of $\mathfrak{su}(n)$ where $n$ $U(1)$ gauge groups connected to form a necklace. Each of the gauge groups is balanced because the balancing condition for a $U(n)$ gauge group is having 2$n$ hypers connected to it. However, for a flavorless quiver with \emph{only} unitary gauge groups, there is an overall $U(1)$ factor that decouples. From the brane setup point of view, this is to fix the center of mass of the brane system. This implies that one of the $U(1)$ gauge groups becomes a $U(1)$ flavor group. As shown by Eq.\eqref{dimformula}, decoupling a $U(1)$ gauge group increases the quaternionic dimension of the Higgs branch from 0 to 1. The quaternionic dimension of the necklace quiver Higgs branch is $\text{dim}_{\mathbb{H}}=1$, corresponding to the orbifold $\mathbb{H}/\mathbb{Z}_n$. Because of this decoupling of $U(1)$ for flavorless unitary quivers, a balanced unitary quiver will never have a trivial Higgs branch.

If a $U(n)$ gauge group is connected to $2n-1$ hypers, it is called \emph{ugly}. Unitary quivers with ugly nodes can have a trivial Higgs branch dimension but will be free theories. A $U(n)$ gauge group with fewer than $2n-1$ hypers is considered \emph{bad}. Bad quivers often lead to incomplete Higgsing (where gauge groups are not fully broken by assigning VEVs to scalars), allowing for a trivial Higgs branch. However, the Coulomb branch of bad quivers exhibits several undesirable properties, such as failing to form a hyperKähler cone \cite{Assel:2017jgo}, a mismatch in the conformal dimension of UV monopoles in the IR, resulting in the failure of the monopole formula, and the breakdown of 3d mirror symmetry \cite{USU}. To avoid these undesirable traits, we focus on finding examples in which all gauge groups are good.

For a good quiver gauge theory to have a trivial Higgs branch, it needs to be balanced and flavorless, which can be achieved if there is at least one special unitary, special orthogonal, or symplectic gauge group. From the brane setup perspective, this is because these theories have their center of mass already fixed (e.g., by an orientifold plane). Therefore, there is no need for decoupling an overall $U(1)$. A similar construction can be done by taking $T[SU(2n)]$ and gauging the $SU(n) \times SU(n) \subset SU(2n)$ flavor symmetry. Using the convention that the balancing condition for an $SU(n)$ gauge group is $2n-1$ hypers, this results in a fully balanced quiver. However, explicit computation using the monopole formula shows that the Coulomb branch Hilbert series diverges. This behavior is also observed in certain fully balanced orthosymplectic quivers in an upcoming work \cite{fully}.
It seems that sometimes the Coulomb branch Hilbert series diverges despite all gauge nodes being good. This is an intriguing phenomenon for which we do not yet have an explanation. In some other examples of fully balanced orthosymplectic quivers, we observe a free theory, consistent with the proposal of a trivial Higgs branch. Quiver \eqref{generalfamily} remains the only example of a fully balanced quiver with a convergent Hilbert series that is not a free theory. If quiver \eqref{generalfamily} had not been found, it might suggest a conspiracy against the existence of a (completely Higgsable) interacting theory with a trivial Higgs branch but non-trivial Coulomb branch.

Since quiver \eqref{generalfamily} is self-dual, its 3d mirror can be obtained by taking $T[SO(2n)]$ and gauging the $SO(n) \times SO(n) \subset SO(2n)$ topological symmetry.

\section{Implications}\label{THEimplications}
If one is satisfied with \emph{any} 3d $\mathcal{N}=4$ theory with a trivial Higgs branch and a non-trivial Coulomb branch, it’s not too difficult to find such examples when scanning the SCFT landscape. However, almost all of these will be bad quivers with incomplete Higgsings and have Coulomb branches that aren’t hyperKähler cones. These properties make it impossible to study the moduli space using efficient tools like the  monopole formula and 3d superconformal index. Furthermore, bad quivers lacking a hyperKähler cone Coulomb branch also prevent the use of 3d mirror symmetry and are ineligible as magnetic quivers for higher-dimensional theories.


In this paper, we have shown that it’s possible to get Lagrangian 3d $\mathcal{N}=4$ theories without these issues and how to systematically find more of them. With the ideal properties of complete Higgsing and hyperK\"ahler cone moduli spaces, we now discuss their implications for $\text{rank}\scriptsize{-}0$ 4d SCFTs and 3d symplectic duality.

\subsection{Relation to 4d $\mathcal{N}=2$ SCFT}
The mirrors of quiver \eqref{generalfamily} have trivial Coulomb branches and non-trivial Higgs branches. The elephant in the room is whether one can uplift these theories to 4d and say something about $\text{rank}\scriptsize{-}0$ $4d$ $\mathcal{N}=2$ SCFTs. The existence of such theories will have significant impact on the classification of 4d SCFTs by the Coulomb branch dimensions \cite{Argyres:2015ffa,
 Argyres:2015gha,
 Argyres:2016xua}. Since the mirrors of \eqref{generalfamily} are not free theories nor discrete gaugings thereof, it is likely that the uplift will also be interacting SCFTs. Most arguments against the existence of $\text{rank}\scriptsize{-}0$ SCFTs in 4d stem from the fact that they haven’t yet been found in the literature. One takeaway of this paper is that a quiver like \eqref{quiver} that has all the ideal properties emerges in a highly non-trivial way. In 3d, it is much easier to construct SCFTs, as most have a UV Lagrangian quiver description. Nevertheless, only after the development of many new methods was a systematic search for exotic types of orthosymplectic quivers possible, and these are the simplest examples found. Thus, a $\text{rank}\scriptsize{-}0$ 4d SCFT might exist in a corner of the landscape, among more complicated and unexplored theories. 

A recent paper \cite{Elliot:2024hat} examined a quiver closely related to \eqref{quiver}. The only difference is the presence of an additional $Sp(1)$ gauge node attached to one of the two $SO(4)$ nodes at the bifurcation. Let us postulate that these two orthosymplectic quivers are magnetic quivers for some $4d$ $\mathcal{N}=2$ theory—specifically, that the Coulomb branch of the orthosymplectic quivers corresponds to the Higgs branch of the 4d theories. Applying an orthosymplectic version \cite{Bao:2024eoq,ospdecay} of the Decay and Fission algorithm \cite{Decay,Fission}, we observe that the quiver in \cite{Elliot:2024hat} decays into \eqref{quiver}, implying that there is a Higgsing relation between their corresponding 4d theories.
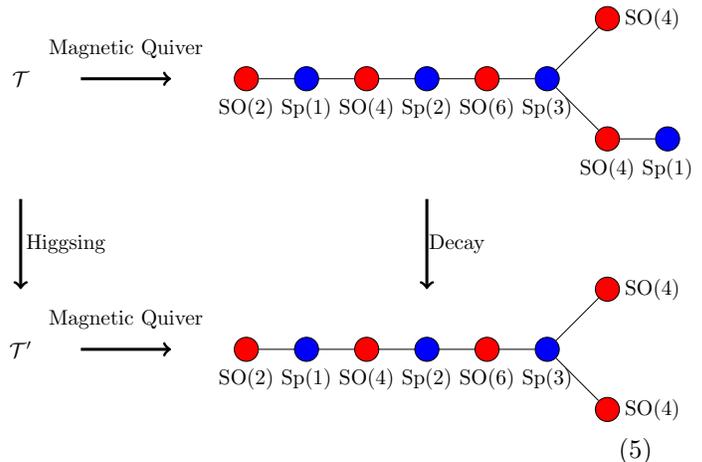
\begin{equation}
    \scalebox{0.8}{\begin{tikzpicture}
	\begin{pgfonlayer}{nodelayer}
		\node [style=redgauge] (0) at (-3, -1) {};
		\node [style=redgauge] (1) at (-1, -1) {};
		\node [style=redgauge] (2) at (1, -1) {};
		\node [style=redgauge] (3) at (3, 0) {};
		\node [style=redgauge] (4) at (3, -2) {};
		\node [style=bluegauge] (5) at (-2, -1) {};
		\node [style=bluegauge] (6) at (0, -1) {};
		\node [style=bluegauge] (7) at (2, -1) {};
		\node [style=none] (8) at (-3, -1.5) {SO(2)};
		\node [style=none] (9) at (-2, -1.5) {Sp(1)};
		\node [style=none] (10) at (-1, -1.5) {SO(4)};
		\node [style=none] (11) at (0, -1.5) {Sp(2)};
		\node [style=none] (12) at (1, -1.5) {SO(6)};
		\node [style=none] (13) at (2, -1.5) {Sp(3)};
		\node [style=none] (14) at (3.75, 0) {SO(4)};
		\node [style=none] (15) at (3.75, -2) {SO(4)};
		\node [style=redgauge] (16) at (-3, 3.5) {};
		\node [style=redgauge] (17) at (-1, 3.5) {};
		\node [style=redgauge] (18) at (1, 3.5) {};
		\node [style=redgauge] (19) at (3, 4.5) {};
		\node [style=redgauge] (20) at (3, 2.5) {};
		\node [style=bluegauge] (21) at (-2, 3.5) {};
		\node [style=bluegauge] (22) at (0, 3.5) {};
		\node [style=bluegauge] (23) at (2, 3.5) {};
		\node [style=none] (24) at (-3, 3) {SO(2)};
		\node [style=none] (25) at (-2, 3) {Sp(1)};
		\node [style=none] (26) at (-1, 3) {SO(4)};
		\node [style=none] (27) at (0, 3) {Sp(2)};
		\node [style=none] (28) at (1, 3) {SO(6)};
		\node [style=none] (29) at (2, 3) {Sp(3)};
		\node [style=none] (30) at (3.75, 4.5) {SO(4)};
		\node [style=none] (31) at (3, 2) {SO(4)};
		\node [style=bluegauge] (32) at (4, 2.5) {};
		\node [style=none] (33) at (4, 2) {Sp(1)};
		\node [style=none] (34) at (0, 1.5) {};
		\node [style=none] (35) at (0, 0) {};
		\node [style=none] (36) at (0.5, 0.75) {Decay};
		\node [style=none] (37) at (-5.75, 3.5) {};
		\node [style=none] (38) at (-4.25, 3.5) {};
		\node [style=none] (39) at (-5, 4) {Magnetic Quiver};
		\node [style=none] (40) at (-5.75, -1) {};
		\node [style=none] (41) at (-4.25, -1) {};
		\node [style=none] (42) at (-5, -0.5) {Magnetic Quiver};
		\node [style=none] (43) at (-6.75, 3.5) {$\mathcal{T}$};
		\node [style=none] (44) at (-6.75, -1) {$\mathcal{T}'$};
		\node [style=none] (45) at (-6.75, 1.5) {};
		\node [style=none] (46) at (-6.75, 0) {};
		\node [style=none] (47) at (-6, 0.75) {Higgsing};
	\end{pgfonlayer}
	\begin{pgfonlayer}{edgelayer}
		\draw (0) to (7);
		\draw (7) to (3);
		\draw (4) to (7);
		\draw (16) to (23);
		\draw (23) to (19);
		\draw (20) to (23);
		\draw (20) to (32);
		\draw [style=arrowed] (34.center) to (35.center);
		\draw [style=arrowed] (37.center) to (38.center);
		\draw [style=arrowed] (40.center) to (41.center);
		\draw [style=arrowed] (45.center) to (46.center);
	\end{pgfonlayer}
\end{tikzpicture}
}
\end{equation}
The Coulomb branch of the top orthosymplectic quiver in \cite{Elliot:2024hat} is described as the two-$F_4$ instanton moduli space. Therefore, the Higgsing to a moduli space of $\mathrm{(one-} F_4 ; \mathrm{instanton})^2$ is expected. Since the Higgsing can also be achieved via nilpotent Higgsing of class $\mathcal{S}$ fixtures, \cite{Elliot:2024hat} proposed a candidate 4d theory for \eqref{quiver} as a class $\mathcal{S}$ vertex operator algebra (VOA) with a twisted non-simply laced $B_3$ algebra. The class $\mathcal{S}$ fixture is characterized by the partitions $([1^7],[3^2,1],[3^2,1])$, which label the Young tableaux of the punctures. This is an unusual case, as these are not the typical class $\mathcal{S}$ theories associated with ADE algebras, and their existence as 4d SCFTs is still under investigation. While \cite{Elliot:2024hat} raised some arguments against their existence, their status remains unclear. The conjectured existence of $4d$ $\mathcal{N}=2$ SCFTs with $F_4$ flavor symmetries and Higgs branches related to instanton moduli spaces has been a long-standing problem. Some of the arguments supporting their existence come from the VOA perspective and their appearance in the Deligne–Cvitanović sequence (see discussion in \cite{Beem:2019snk}).


The generalization to the 4d uplifted mirrors of quiver \eqref{generalfamily} will then be the $B_{n-1}$ VOA with $([1^{2n-1}],[(n-1)^2,1],[(n-1)^2,1])$.  

 It is difficult to conduct a systematic search for rank-zero 4d SCFTs because one does not know how to even begin. However, finding 3d $\mathcal{N}=4$ theories with trivial Coulomb branches is possible owing to the power of 3d mirror symmetry. In an upcoming work \cite{fully}, we construct all possible orthosymplectic quivers that are fully balanced. It turns out there are only a finite number of such families, as the balancing conditions are highly constraining. As discussed above, all these balanced quivers have trivial Higgs branches, and their 3d mirrors will have trivial Coulomb branches. This make them possible candidates for rank-zero $4d$ $\mathcal{N}=2$ SCFTs compactified on a circle. Furthermore, many of these orthosymplectic quivers take the form of star-shaped quivers, which are ideal candidates for being magnetic quivers of class $\mathcal{S}$ theories. This approach enables a more systematic search for rank-zero SCFTs. Ultimately, if none of these 3d theories can be matched with a 4d SCFT, it could provide insights into the absence of 4d rank-zero theories. 


\subsection{Symplectic duality}

In the landscape of good quiver gauge theories found in the literature, it is often challenging to find two theories that share the same Coulomb (Higgs) branch but have different Higgs (Coulomb) branches. This leads to attempts to use symplectic duality to relate the Coulomb branch to the Higgs branch of the \emph{same} theory \cite{Bullimore:2016nji}. Our quivers \eqref{generalfamily} satisfy the condition that both branches are always hyperKähler cones, making them good candidates for performing symplectic duality \cite{braden2022quantizationsconicalsymplecticresolutions}. However, an immediate issue arises as not only do we have infinitely many non-identical Coulomb branches mapping into the same Higgs branch, but the Higgs branch is also trivial (a point). This surjection demonstrates that, even with complete knowledge of the (trivial) Higgs branch, one cannot infer anything about the Coulomb branches of these theories. This interesting example suggests that, even among quiver gauge theories where both moduli spaces are singular hyperKähler cones, there appears to be a limitation on the types of $3d$ $\mathcal{N}=4$ SCFTs that symplectic duality can act on.




\medskip\noindent {\bf Acknowledgements} ZZ deeply appreciates discussions with Philip Argyres, Christopher Beem,  Antoine Bourget, Nicklas Garner, Amihay Hanany, Heeyeon Kim, Kimyeong Lee, Matteo Sacchi, Sakura Sch\"afer-Nameki,  and Ben Webster. ZZ is supported by the ERC Consolidator Grant \# 864828 ``Algebraic Foundations of Supersymmetric Quantum Field Theory'' (SCFTAlg).  We would also like to thank the Pollica Physics Centre for hospitality while some of this
 work was completed.

\bibliographystyle{apsrev4-1}
\bibliography{bibli}
\end{document}